# Alphabet-dependent Parallel Algorithm for Suffix Tree Construction for pattern searching

Freeson Kaniwa, Venu Madhav Kuthadi, Otlhapile Dinakenyane
and Heiko Schroeder

*Botswana International University of Science and Technology*
*fkaniwa@gmail.com*

*Abstract*

*Suffix trees have recently become very successful data structures in handling large data sequences such as DNA or Protein sequences. Consequently parallel architectures have become ubiquitous. We present a novel alphabet-dependent parallel algorithm which attempts to take advantage of the perverseness of the multicore architecture. Microsatellites are important for their biological relevance hence our algorithm is based on time efficient construction for identification of such. We experimentally achieved up to 15x speedup over the sequential algorithm on different input sizes of biological sequences.*

## 1. Introduction

There has been a technological breakthrough in the last decade, which saw a huge improvement in DNA sequencing technologies [1-3] and availability of multicore architectures in desktop computers [4-5].

The DNA sequences are very large and can span to 3 billion characters for the whole human genome. Such sizes have performance issues during analysis hence this attempt to improve the analysis by taking advantage of multicore architectures which has become ubiquitous. One alternate step in performance challenge is to take advantage of the multicore revolution. To do this, algorithms and data structures should be modified to make them well suited for the parallel architectures [4]. Most of these tools for analysis of genomes are sequential, memory intensive and therefore cannot easily scale to larger genomes [2]. One such data structure is the suffix tree. The suffix tree was first introduced by Weiner. A suffix tree of a string *s* represents all suffices of *S*. A suffix tree is a very useful data structure with many applications and many sequential algorithms for suffix tree construction have been proposed [6-9]. However all these suffer from high performance requirements such as in string processing algorithms which processes huge sequences like DNA or Protein sequences. Due to this challenge, there is need for new approaches to improve performance on this data structure [10, 11]. A very strong property of a suffix tree is that, after preprocessing and construction, a pattern, *p*, can be found in a string, *S*, in O ($|p|$), furthermore in O ($|p|+k$) where *k*, is the number of occurrences of *p*.

Generally, when considering a parallel solution for suffix tree construction, the most difficult part, is on partitioning the tree [12]. In this paper we attempt to provide an In-memory parallel suffix tree construction approach to improve the time requirements. We partition the tree basing on the alphabet of the string which yields independent construction. This paper proposes a novel alphabet-dependant parallel algorithm for suffix tree construction (kmer-based) involving assigning suffices to processors based on their initial character. The algorithm achieves a performance speedup of up to 15x over the sequential algorithm.

The remainder of the paper is organized as follows; we introduce the Preliminaries, that is, the foundation of the concepts used in the entire paper in Section 2. We provide brief related work in Section 3 and provide our parallel algorithm in section 4 and then experiments and analysis of results in Section 5. The paper is concluded with a conclusion section in Section 6 which discusses the summary and future work.

## 2. Preliminaries

Analysis of our algorithm is done using the Circuit Model of the Work Depth model, which uses Concurrent Read and Exclusive Write (CREW). In this model, the *depth* D is the number of time steps needed and *work* W is the number of operations needed. if $P$ processors are available, using Brent's scheduling theorem [24] ,we can calculate the running time as O ($T_w$ /P + $T_d$) [14]. Then the *speedup* of the algorithm will be equal to O ($T_1/T_\infty$) where $T_1$ is equal to time under 1 processor and $T_\infty$ time under multiple processors. Concurrent reading and exclusive writing is allowed in our model. This model can still be converted to PRAM model by using the Brent Theorem as stated earlier [13].

In order to show parallel execution, we will include three concurrency keywords, that is, *spawn*, *sync* and *parfor* as done in [13] The concurrency keywords indicates which parts of computation should be executed in parallel thus expressing logical parallelism, The *spawn* keyword represents that the procedure call that comes before it will be executed in parallel. The *sync* keyword shows that all spawned procedures must complete before going to the next instruction in the stream. Lastly, *parfor* which we will use, is a parallel for loop which is the same as the ordinary for loop in serial algorithms, except that the iterations can execute concurrently. We denote a sequence of length *n* by $S = S[0],…,S[n-1]\}$, where $S[i]$ is the *i*'th symbol of $S$. We also denote an alphabet by $\sum = \{0,…,\sigma – 1\}$[10], where σ is the alphabet size. We use |p/ to represent, *n* is the length of the string $S$, σ is the size of the alphabet and $\sum$ is the set of the alphabet.

Although the programming community has adopted the random access model for serial algorithms, no single model for parallel algorithms has gained wide acceptance. This is the case since parallel computers can have shared memory model or distributed memory model hence vendors have not agreed on a single architectural model. However each new desktop and laptop is now a shared memory parallel computer and this trend appears to be going towards the shared memory multiprocessing. Although time will tell, in this paper we will also take such approach or route.

Let $S = s_1…s_n$ be a string with some symbols of the alphabet $\sum$ . In this paper the suffix tree of string *S*, is a rooted tree with edges and nodes that are labeled with their indices thereby using constant space for each label. The suffix tree will satisfy the following properties:
Property 1:
1. The root edges store the indices of different characters that starts with different symbols away from the root.
2. Every internal node has at least two children.
3. Each node is labeled by the index positions formed by combining edge labels from root to the node.

An example of a suffix tree for string $S = \{xabxac\}$ with σ = 4, numbers on the leaves represent indexes of the suffices as shown in Figure. 1. At this this point the suffix tree is an implicit suffix tree, since a suffix can be a prefix of another suffix.

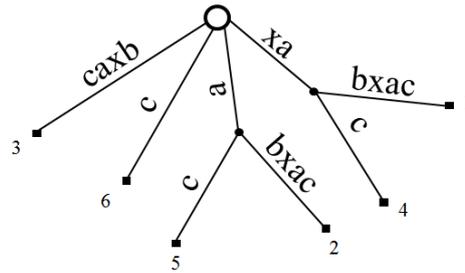

**Figure 1. Implicit Suffix Tree of S**

It is a standard practice to append the string with '$' as the terminal symbol so as to generate an explicit suffix tree, as shown in Figure. 2. With suffices for S = {*xabxac$*, *abxac$*, *bxac$*, *xac$*, *ac$*, *c$*, $}

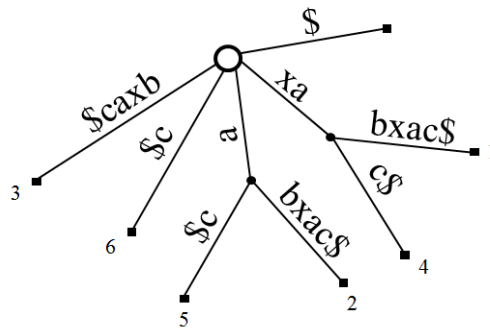

**Figure 2. Explicit Suffix Tree of S**

## 3. Related work

In this section, we briefly survey the parallel algorithms for construction of suffix trees. Some work related to parallel processing of suffix trees has been done. There are basically two classes of suffix tree algorithms, that is, In-memory and Disk-based. In-memory can be used if both the string *S* and the suffix tree can fit in the main memory and these classes of algorithms suffer from poor locality reference. However the latter tries to address this problem by storing some parts of the string *S* or suffix tree on the disk memory.

The first O (*n*) sequential algorithm for suffix tree construction was done by Weiner [15]. The algorithm takes O (*m* log σ) time, where σ =|∑|, the number of characters of the alphabet. McCreight [9] improved this algorithm by giving it a more efficient construction within the same asymptotic bound [7].

Apostolico et. al. [16] proposed the first parallel algorithm for suffix tree construction. This parallel algorithm was based on doubling approach with a technique they called *naming*. Given a parameter value 0 < ϵ ≤ 1, the algorithm runs in O ($\frac{1}{v} \log n$) time, O ($\frac{n}{\epsilon} \log n$) work and $O(n^{1+\epsilon})$ space using CRCW PRAM Model. On binary alphabet, this algorithm is not very space efficient since it has polynomial space complexity. Hariharan [17] later improved the parallel construction of suffix trees to linear space and time. The algorithm takes O (log$^4$ *n*) time and O (*n*) for work and space on CREW PRAM model. A further improvement was also done by Farach and Muthukrishnan to O (log *n*) time on randomized CRCW PRAM [29]. Very recently Satish et al. [2] proposed a parallel algorithm for suffix tree construction. Their parallel algorithm is based on the MapReduce programming model. Their approach *vertically partitions* the suffix tree and then constructs the subtrees independently in parallel by extending Ukkonnen's algorithm [6] as shown in Fig. 3.

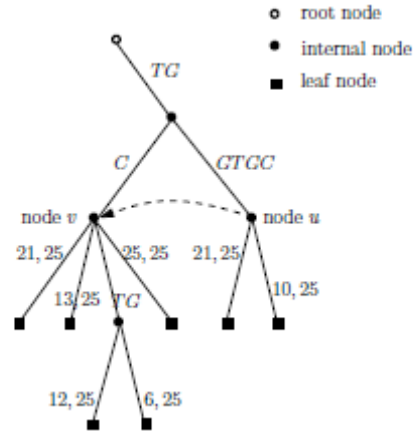

**Figure 3. Sub-tree for S-prefix T and sample suffix link in its context [2]**

Currently there are only two parallel practical implementation of the suffix tree construction, that is, Wavefront [19] and ERA [20]. Hence this paper has emphasis on parallel implementation of the parallel algorithm. The Wavefront works by splitting the sequence into partitions of the resulting complete tree. The partitioning process is done by making use of variable length prefixes making sure that every partition starts with the same prefix thereby guaranteeing independence of each sub tree which can then later be merged. ERA also partitions the sequence into independent sub-trees which are then subdivided into partitions making it possible to process each partition in memory. Also most recently Comin and Ferrares [21] introduced their parallel suffix tree algorithm called Parallel Continuous Flow (PCF) which works by dividing a string into independent sub trees with a time complexity of $O(log^2(N/\sqrt{P})$. These sub trees are further divided into partitions which then can be processed in memory and works as shown in Fig. 4:

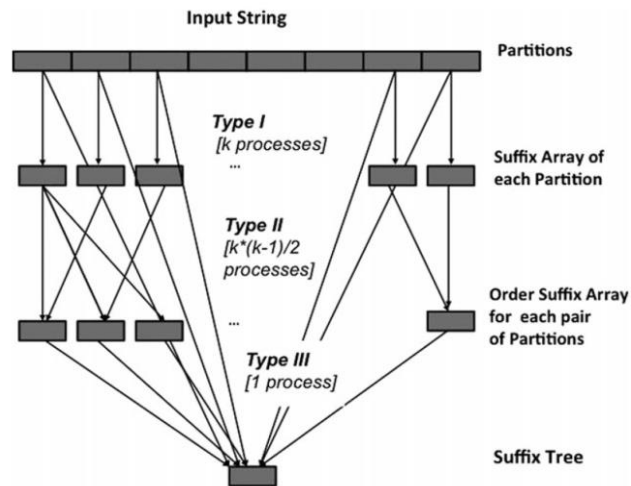

**Figure 4. A workflow of data dependencies [21]**

## 4. Parallel Suffix Tree Construction

We now describe our parallel algorithm for suffix tree construction with the help of an example. Our algorithm adopts a Generalized Suffix Tree idea and constructs the suffix tree, branch by branch in parallel and each branch is a sub-tree which will later be merged to form the complete suffix tree. We make use of Example 1. to explain our idea.

**Example 1**: Given a string $S = \{aaabaca,…\}$ the algorithm first extract the alphabet from $S$, $\sum = \{a,b,c\}$, each processor processes all suffices starting with the same symbol of the alphabet and these are executed independently in parallel. For example, given the String *S* = {*acgttacg...*}, and we need to search for all repeats of size 4 then the first step is to extracts the alphabet ∑ = {*a,c,g,t,...,λ*}(where λ is the last symbol in the sequence) and generate the sets ($P_α$ where *P* is the set containing all the suffices starting the symbol α) by grouping all the suffices which start with same symbol of the alphabet with exact lengths of 4, that is,

   1$^{st}$ Phase:
   $P_a$ = {*acgt*}  and {acg} ∉ $P_a$

   $P_c$ = {*cgtt*} and {cg} ∉ $P_c$

   $P_g$ = {*gtta*} and {g} ∉ $P_g$

   $P_t$ = {*ttac, tacg*}

   $P_n$ = { λ ..., λ ...}

To create the suffix tree, we introduce a terminal symbol, $, (as stated in section 2), to the substrings hence suffices. We do so by following the following two rules which will guarantee an explicit suffix tree. And this leads us to the final phase.

   1. No suffix of *S* is a prefix of a different suffix of *S*.
   2. There is a leaf for each suffix of *S*.

Final Phase:

   $P_a$ = { *acgt$*}

   $P_c$ = {*cgtt$*}

   $P_g$ = {*gtta$*}

   $P_t$ = {*ttac$, tacg$*}

   $P_n$ = { λ ...$, λ ...$,...}

Our algorithm then construct the suffix tree branch wise independently for each set in parallel as shown in Fig. 5. The construction of branches is done in parallel however the construction of the generalized sub suffix tree for each branch is done sequentially. That is, the branches are computed sequentially in an ordinary generalized suffix tree fashion word by word but using the sliding window approach. Finally the algorithm merges these branches to a common root to come up with the complete suffix tree for the string *S* as shown in Fig. 6. The merging process is trivial since they all just need the common root to generate the complete suffix tree which can reduce the cost of expensive I/O.

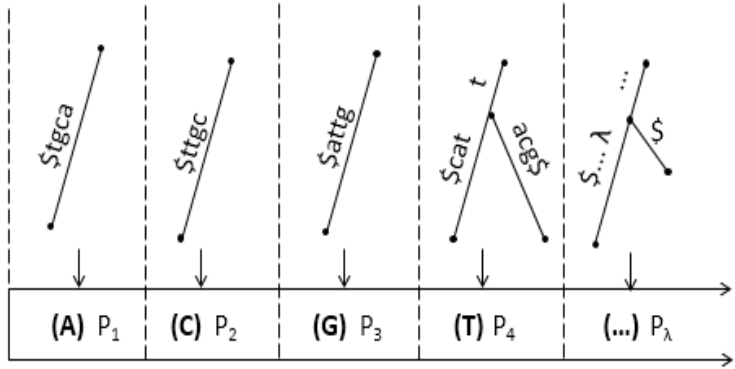

**Figure 5. Branch-wise Parallel Construction of Suffix Tree of S**

Our algorithm is based from the observation that all the root edges of a suffix tree are derived from the alphabet in which each character in the alphabet maps to a root edge in the tree.

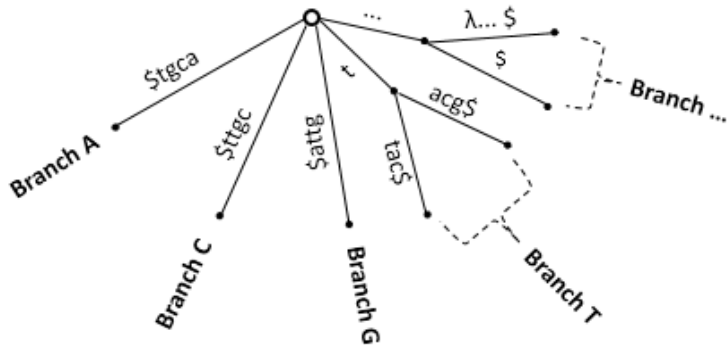

**Figure 6. Merged Complete Suffix Tree (Example 1)**

Lemma 1: If the size of the alphabet of a string $S$ is $\sigma = |\Sigma|$) then the number of the root edges for the Suffix Tree, $T$ of $S = \sigma + 1$, given that we have used all the alphabet characters.

Proof: Let $S = abaabc$, then $\Sigma = \{a,b,c\}$ and $\sigma = 3$ and the required pattern is of size 3 ($|p|$). Thus we assume that $e \neq 4$ where $e$ is the number of root edges of the suffix tree of S. So we then assume that, the suffix tree has $e = 5$. We want to prove that, $e = \sigma + 1$ given that $\sigma = 3$. We construct the suffix tree for $S$ in Figure 7. $S$ has the following set of $k$-mer substrings $\{aba, baa, aab, abc\}$, by property 1(Section 2) which states that "*The root edges stores the indices of different characters that starts with different symbols away from the root*" To construct the suffix tree, we now introduce $ as our terminal symbol to maintain the definition of an implicit suffix tree where $\$ < \Sigma$ and $\$ \notin \Sigma$. So finally we have $\{aba\$, baa\$, aab\$, abc\$, \$\}$ as our substrings. We show the suffix tree of $S$ in Figure. 7.

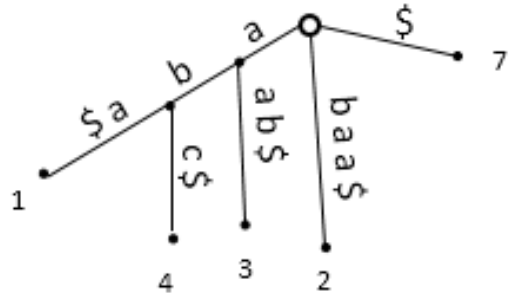

**Figure 7. Suffix Tree of {abaabc} with |p| =3**

As shown in Fig. 7, $e \neq 5$ as assumed earlier but rather $e = 4$, that is, $e = \sigma + 1$ if all the alphabet characters have been used.

□

We present our parallel algorithm in Figure 7. as pseudo-code and provide follow-up description of our algorithm.

```
Algorithm 1. Parallel algorithm for Suffix Tree construction (PaST)
Input   : S, n, ∑,|p|  // S – string, n – number of characters, ∑ - Alphabet
          // |p| - pattern size or repeat size
Output: A suffix tree representation ST of S

1 A [∑ = {0,..., σ – 1}] := {} // set of alphabet characters

2 σ :=|∑| // size of the alphabet

3 ST_b := ST_c :={}
   // initialize the ST_b (Suffix Tree branch) and ST_c (Complete Suffix Tree)

4 parfor i ← 0 to σ – 1 do

5     for j ← 0 to n-1 do

6         if (A[i] = S[j]) then

7             spawn ST_{b_A[i]} ← S [j... j + (|p| – 1)]
              // ST_{b_A[i]} construct subtree with respect to the alphabet character
8 sync

9 for i ← 0 to σ – 1 do

10    ST_c ∈ ST_{b_A[i]} //Merge the branches to complete suffix tree

11 return ST_c
```

**Figure 8: Parallel Algorithm for Suffix Tree construction**

We now describe our suffix tree construction algorithm, expanding from example 1, which constructs all branches of the suffix tree in parallel. We refer to this algorithm as *PaST*. Since each branch needs only suffices starting with the same characters to do the construction, groups of sets of suffices are generated. We made use of the observation from Lemma 1 and Property 1.

We start by extracting all the suffices, that is, those with initial similar symbols and group them in their respective sets (lines 4 - 6) and this extraction is done in parallel and independently since each processor is looking for its symbol in *S*. However there is access control on the shared string/sequence. The second phase starts with the independent construction of branches or sub-trees by taking each suffix to build the respective branches suffix by suffix in (lines 7 - 8) which is similar the process of constructing a generalized suffix tree. Then finally the merging process is done sequentially to come up with a complete suffix tree (lines 9 - 10).

## 5. Experiments & Analysis of Results

Implementation was done in Python. All parts of the algorithm were implemented. We utilized pythons' variant of Mark Carlson C++ code for similar analysis which depends on the Ukkonen O (*n*) time algorithm which is suffix tree based (ST-based) [22], [23].

A. *Experimental setup*

Implementation was done utilizing Python version 3.5.2 and Anaconda version 4.1.1 to our proposed algorithm. The experiments were run on a HP Z820 Workstation with a 3.30 GHz 16-core Intel Xeon E5-2623 chip which has 10 MB L3 Cache, and we applied parallelism on a shared-memory as in our algorithm. The computer runs Ubuntu 14.04LTS (64-bit) OS with 32 GB internal memory, swap space of 64 GB, and 1 TB Serial ATA Hard Disk Drive at 7,200 RPM. We made use of real genomic data of DNA and Protein sequences.

1. The Entire Human Genome sourced from NCBI [24].
2. The X- Chromosome (Human) [24].
3. Protein sequences from Pizza&Chili Corpus [25].

We measure the programs' execution time for the construction only. We use pythons' *time* method to measure the running time of the suffix tree construction process which is measured in seconds as shown in Table 1.
and Table 2.

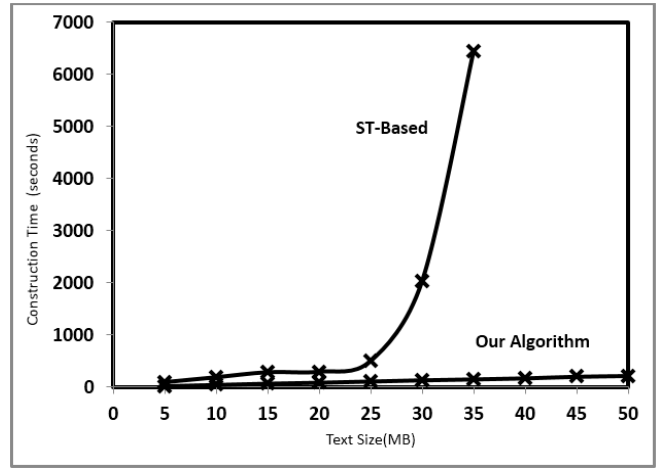

**Figure 9. Construction time comparison on DNA sequences for pattern size of 5**

**Figure 9.** shows construction times of our method versus the ST-based method. Our method was run with specification of pattern size 5, that is, *k* = 5.

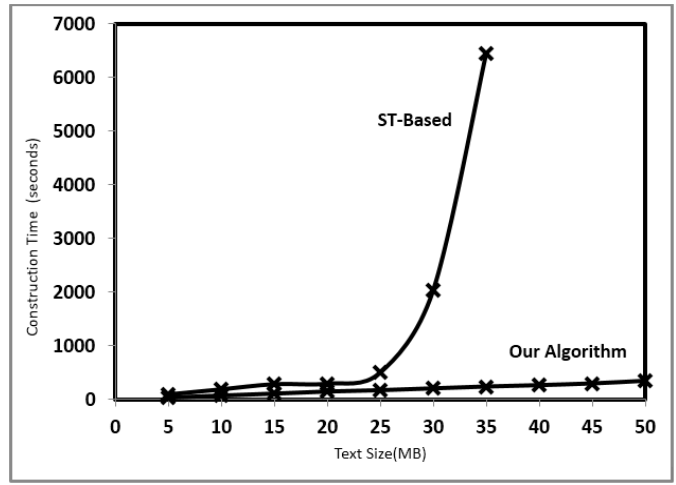

**Figure 10. Construction time comparison on DNA sequences for pattern size of 10**

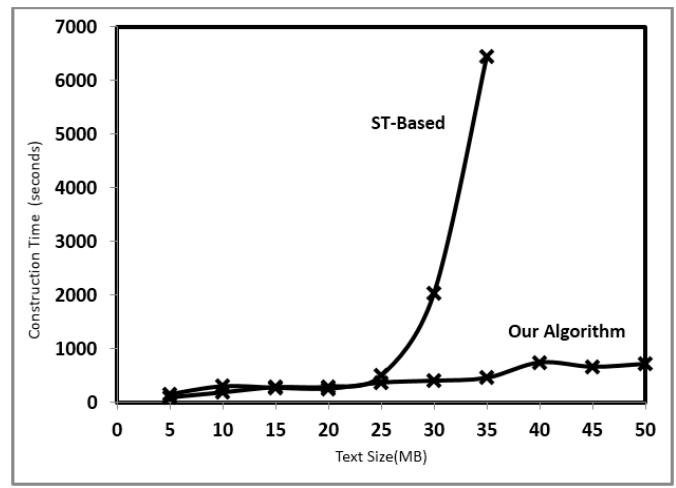

**Figure 11. Construction time comparison on DNA sequences for pattern size of 15**

Table 1. shows the construction times measured in seconds to construct the complete suffix tree per $k = 5, 10$ and $15$. Text size column shows the input sizes in megabytes which varies from 5 upto 50 megabytes.

**Table 1. Construction times of DNA sequences of our algorithm vs. ST-Based for various pattern sizes**

| Text Size (MB) | ST-Based (s) | Our algorithm (s) | | |
|---|---|---|---|---|
| | | k = 5 | k = 10 | k = 15 |
| 5 | 92 | 24 | 35 | 155 |
| 10 | 186 | 48 | 72 | 304 |
| 15 | 283 | 73 | 107 | 277 |
| 20 | 292 | 90 | 146 | 260 |
| 25 | 497 | 113 | 171 | 373 |
| 30 | 2028 | 136 | 204 | 409 |
| 35 | 6444 | 153 | 238 | 462 |
| 40 | n/a | 173 | 263 | 749 |
| 45 | n/a | 203 | 294 | 669 |
| 50 | n/a | 218 | 343 | 722 |

Table 2. depicts construction times for different biological datasets, that is, the full chromosomes and the whole human genome, however the ST-Based did not run to completion for all these datasets even after several days, which we eventually terminated, and this is also discussed in detail under observations section.

**Table 2. Construction times of our algorithm for chromosomes and the complete human genome**

| Dataset | Text Size (MB) | ST-Based (s) | Our algorithm (s) |
|---|---|---|---|
| | | | k = 5 |
| X-Chromosome | ≈ 150 | n/a | 678.180 |
| Chromosome 1 | ≈ 226 | n/a | 1 054.110 |
| Chromosome 2 | ≈ 235 | n/a | 1 088.860 |
| Chromosome 3 | ≈ 193 | n/a | 862.330 |
| Complete Human Genome | ≈ 3000 | n/a | 12 944.460 |

B. *Analysis*

We observed and recorded the running times of our algorithm versus the Suffix Tree-based technique (ST-Based) [22]. The accompanying primary perceptions were noted from this test study:

1. Our algorithm shows improved performance in terms of execution time. The algorithm achieves a performance speedup of up to 15x over the sequential algorithm with better speedup in lower pattern size construction as opposed to bigger pattern size, that is, $k = 5$ versus $k = 15$ (Fig. 12)

2. When the text size becomes larger than 25MB our method shows a huge improvement in terms of time cost compared to the ST-based method. The ST-based time cost increases as the text file exceeds 25MB. Both algorithms perform the same in terms of the running time when the text size if less than 25MB. This improvement can be attributed to parallelism since parallelism benefits are significant when the amount of processing increases.

3. We also observed that the ST-Based method construction does not complete when the text size exceeds 35MB which is possibly due to the high memory requirements of the ST-based method. Our method only requires to construct a tree of equal branch sizes of 5,10 or 15 meaning it has a small memory foot print compared to constructing the complete suffix tree of size *n* where *n* is the size of the string to be indexed.

4. We further noticed that, the construction time of the chromosomes and the whole human genome was impressive since the ST-based could not complete the construction even after several days. The whole human genome of repeat size 5 took approximately 3 and half hours to complete construction. (Table. 2).

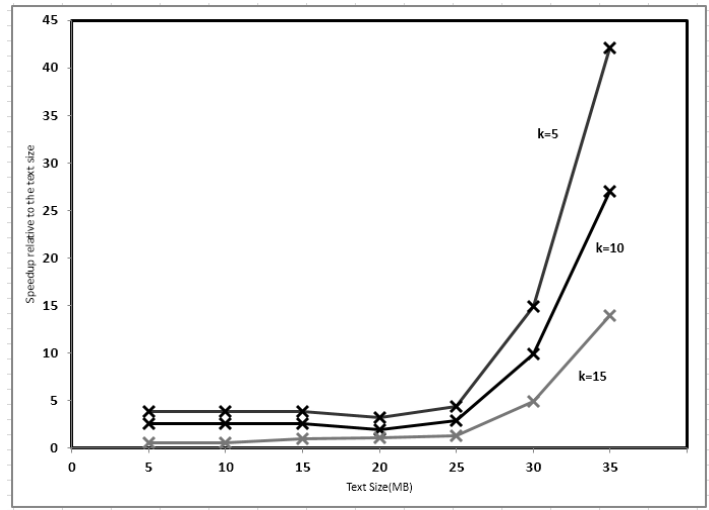

**Figure 12. Speedup relative to text size**

Our algorithm can be useful in detection of small repeats such as microsatellites since they are of size 2 to 6 long. Microsatellites are important since they used as DNA markers for different biological functions[37]. These types of repeats allow you to specify size before detecting them prior to the analysis. The parallel programs ERA [20] and Wavefront [19] were not compared with our parallel program since these two parallel programs implements a complete construction of a suffix tree (and also not generalized suffix tree) whereas ours implements a parallel algorithm of *k*-mer based of a sequential algorithm for the generalized suffix tree. This means, to the best of our knowledge, this is the first parallel version of *k*-mer based approach on a generalized suffix tree.

## 6. Conclusion

In this paper we have introduced a novel approach to constructing a suffix tree in parallel. We have also shown that the suffix tree of a string can be constructed optimally in parallel provided the size of the pattern is known beforehand. And we have consequently shown that the construction time of our parallel method is better than that of the sequential version with a speedup of up to 15x. However the approach still suffers poor scalability due to the nature of parallelization method used, that is, alphabet-dependant, which can be improved in future.

### Acknowledgments

We would like to thank the IT Department of BIUST for providing the HPZ820 workstation for the test runs.